\documentclass[12pt]{article}
\usepackage{amsmath,amsfonts,latexsym,xspace}

\newcommand{\itm}{\item[$\diamond$]}
\newcommand{\strt}{\parbox{0mm}{\vspace{5ex}}}

\newcommand{\smallspace}{\mskip 2mu minus 1mu}
\newcommand{\tinyspace}{\mskip 1mu}

\newcommand{\hpsr}[2]{\phantom{#2}\makebox[0pt][r]{$#1$}}

\newcommand{\sqrta}{{\textstyle\frac{1}{\sqrt a}}}

\def\goodUpsilon{\Upsilon_1}
\def\badUpsilon{\Upsilon_0}
\def\goodPsi{\Psi_1}
\def\badPsi{\Psi_0}

\def\squareforqed{\hbox{\rlap{$\sqcap$}$\sqcup$}}
\def\qed{\ifmmode\squareforqed\else{\unskip\nobreak\hfil
\penalty50\hskip1em\null\nobreak\hfil\squareforqed
\parfillskip=0pt\finalhyphendemerits=0\endgraf}\fi}

\newtheorem{theorem}{Theorem}
\newtheorem{lemma}[theorem]{Lemma}
\newtheorem{corollary}[theorem]{Corollary}
\newtheorem{definition}[theorem]{Definition}
\newenvironment{algorithm}[1]{\medskip\noindent%
\itemsep0pt\begin{trivlist}\item[]%
{\flushleft\textbf{Algorithm( #1 )}}}%
{\end{trivlist}\medskip}
\newenvironment{proof}{\begin{trivlist}\item[]{\flushleft\bf Proof }}
{\qed\end{trivlist}}

\newcommand{\NP}{\mbox{\bf NP}}
\newcommand{\npc}{\mbox{\NP--com}\-plete}

\DeclareMathSymbol{\itTheta}{\mathalpha}{letters}{"02}

\newcommand{\ket}[1]{\mbox{$| #1 \rangle$}}
\newcommand{\bra}[1]{\mbox{$\langle #1 |$}}
\newcommand{\braket}[1]{\mbox{$\langle #1 | #1 \rangle$}}
\newcommand{\ketbra}[1]{\mbox{$| #1 \rangle\langle #1 |$}}

\newcommand{\integer}{{\mathbb Z}}
\newcommand{\reals}{{\mathbb R}}

\newcommand{\identity}{{\mathbf I}}

\newcommand{\mytheta}{{\theta_a}}
\newcommand{\barmytheta}{{\overline{\theta_a}}}

\newcommand{\algqsearch}{\textup{\textbf{QSearch}}\xspace}
\newcommand{\algestamp}%
{\textup{\textbf{Est\hspace{-0.075em}\_Amp}}\xspace}
\newcommand{\algcount}{\textup{\textbf{Count}}\xspace}
\newcommand{\algapproxcount}%
{\textup{\textbf{Basic\hspace{-0.075em}\_Approx\_Count}}\xspace}
\newcommand{\algexactcount}{\textup{\textbf{Exact\_Count}}\xspace}
\newcommand{\algoptapproxcount}%
{\textup{\textbf{Approx\_Count}}\xspace}

\newcommand{\rblc}[1]{\raisebox{#1}{$\lceil$}}
\newcommand{\rbrc}[1]{\raisebox{#1}{$\rceil$}}

\newcommand{\tildetheta}{{\tilde{\theta}}}

\newcommand{\KS}[2]{\mbox{$\ket{  \mathcal{S}_{#2}(#1)}$}}

\newcommand{\BKS}[3]%
{\mbox{$\langle \mathcal{S}_{#3}(#1) | \mathcal{S}_{#3}(#2) \rangle$}}

\def\mylangle{\texttt <}
\def\myrangle{\texttt >}
\newcommand{\myurl}[1]{\mylangle http:/$\mskip -2.5mu$/{#1}\myrangle}

\newlength{\upceil}
\newlength{\upbceil}
\newcommand{\eps}{\varepsilon}
\newcommand{\oneovereps}{{\frac{1}{\raisebox{0.35ex}{\scriptsize $\eps$}}}}

\title{Quantum Amplitude Amplification\\and Estimation}

\author{~~~Gilles Brassard\,%
\thanks{\,D\'epartement~IRO, Universit\'e de Montr\'eal,
 C.P.~6128, succursale centre-ville, Montr\'eal (Qu\'ebec), Canada H3C~3J7.
 email:~\texttt{brassard}%
\textbf{\char"40}\texttt{iro.umontreal.ca}.
Supported in part by Canada's {\sc Nserc}
and Qu\'ebec's {\sc Fcar}.}~~~
\and ~~~Peter H{\o}yer\,%
\thanks{\,BRICS, Department of Computer Science, University of Aarhus,
Ny~Munkegade, Bldg.~540, DK-8000 Aarhus~C, Denmark.
email:~\mbox{\texttt{hoyer}\textbf{\char"40}\texttt{brics.dk}}.
Part of this work was done while at
D\'epartement IRO, Universit\'e de Montr\'eal.
Basic Research in Computer Science is supported by
the Danish National Research Foundation.}~~~
\and ~~~Michele Mosca\,%
\thanks{\,CACR, 
Department of C\&O, 
Faculty of Mathematics,
University of Waterloo, 
\mbox{Waterloo}, Ontario, Canada \mbox{N2L 3G1}.
email:~\mbox{\texttt{mmosca}\textbf{\char"40}\texttt{cacr.math.uwaterloo.ca}}.
Most of this work was done while at
Centre for Quantum Computation, Clarendon Laboratory,
University of Oxford.
Supported in part by Canada's \textsc{Nserc} and UK's \textsc{Cesg}.}~~~
\and ~~~Alain Tapp\,%
\thanks{\,CACR, 
email:~\mbox{\texttt{atapp}\textbf{\char"40}\texttt{cacr.math.uwaterloo.ca}}.
Most of this work was done while at
D\'epartement IRO, Universit\'e de Montr\'eal.
Supported in part by a postdoctoral fellowship from
Canada's {\sc nserc}.}~~~
}

\date{2 May 2000}

\begin{document}

\maketitle
\thispagestyle{empty}

\begin{abstract}
Consider a Boolean function \mbox{$\chi: X \rightarrow \{0,1\}$} that
partitions set $X$ between its \emph{good} and \emph{bad} elements,
where $x$ is good if \mbox{$\chi(x)=1$} and bad otherwise.
Consider also a quantum algorithm $\mathcal A$
such that \mbox{${\mathcal A}\ket{0} = \sum_{x\in X} \alpha_x \ket{x}$}
is a quantum superposition of the elements of~$X$,
and let $a$ denote the probability that a good element is produced
if ${\mathcal A}\ket{0}$ is measured.
If~we repeat the process of
running $\mathcal A$, measuring the output, and using $\chi$ to check
the validity of the result, we shall expect to repeat $1/a$ times on the
average before a solution is found.  \emph{Amplitude amplification} is
a process that allows to find a good $x$ after an expected number of
applications of ${\mathcal A}$ and its inverse which is proportional
to $1/\sqrt{a}$, assuming algorithm ${\mathcal A}$ makes no measurements.
This is a generalization of Grover's searching algorithm in which
${\mathcal A}$ was restricted to producing an equal superposition
of all members of~$X$ and we had a promise that a single $x$ existed
such that \mbox{$\chi(x)=1$}.  Our algorithm works whether or not the
value of $a$ is known ahead of time.  In case the value of $a$ is known,
we can find a good $x$ after a number of applications of
${\mathcal A}$ and its inverse which is proportional to $1/\sqrt{a}$
even in the worst case. We~show that this quadratic speedup
can also be obtained for a large family of
search problems for which good classical heuristics exist.
Finally, as our main result,
we combine ideas from Grover's and Shor's quantum algorithms
to perform \emph{amplitude estimation}, a process that allows to
estimate the value of $a$.
We~apply amplitude estimation to the problem of \emph{approximate counting},
in which we wish to estimate the number of $x\in X$ such that
\mbox{$\chi(x)=1$}.  We~obtain optimal quantum algorithms in a
variety of settings.
\end{abstract}

Keywords: \emph{
 Quantum computation.
 Searching.
 Counting.
 Lower bound. }


\section{Introduction}

Quantum computing is a field at the junction
of theoretical modern physics and
theoretical computer science.
Practical experiments involving a few quantum bits
have been successfully performed, and
much progress has been achieved
in quantum information theory, quantum error correction
and fault tolerant quantum computation.
Although we are still far from
having desktop quantum computers in our offices,
the quantum computational paradigm
could soon be more than mere theoretical
exercise.

The discovery by Peter Shor~\cite{Shor} of a polynomial-time
quantum algorithm for factoring and computing discrete logarithms
was a major milestone in the history of quantum computing.
Another significant result is Lov Grover's quantum search
algorithm~\cite{Grover1,Grover2}.
Grover's algorithm does not solve \npc{}
problems in polynomial time,
but the wide range of its applications
more than compensates for this.

In this paper, we generalize Grover's algorithm in a variety of directions.
Consider a problem that is characterized by a Boolean function $\chi(x,y)$
in the sense that $y$ is a good solution to instance~$x$ if and only if
\mbox{$\chi(x,y)=1$}.
(There could be more than one good solution to a given instance.)
If~we have a probabilistic algorithm $\mathcal P$ that outputs a
guess ${\mathcal P}(x)$ on input~$x$, we can call $\mathcal P$ and
$\chi$ repeatedly until a solution to instance $x$ is found.
If~$\chi(x,{\mathcal P}(x))=1$ with probability \mbox{$p_x>0$}, we
expect to repeat this process $1/p_x$ times on the average.
Consider now the case when we have a quantum algorithm $\mathcal A$
instead of the probabilistic algorithm.  Assume $\mathcal A$ makes
no measurements: instead of a classical answer, it produces quantum
superposition $\ket{\Psi_x}$ when run on input~$x$.  Let $a_x$
denote the probability that $\ket{\Psi_x}$,
{\em if measured}, would be a good solution.  If~we repeat the process of
running $\mathcal A$ on~$x$, measuring the output, and using $\chi$ to check
the validity of the result, we shall expect to repeat $1/a_x$ times on the
average before a solution is found.  This is no better than the classical
probabilistic paradigm.

In~Section~\ref{sec:ampl}, we describe a more efficient approach to this
problem, which we call \mbox{amplitude amplification}.  Intuitively, the
probabilistic paradigm increases the probability of success roughly by a
constant on each iteration; by contrast, amplitude amplification increases the
{\em amplitude\/} of success roughly by a constant on each iteration.  Because
amplitudes correspond to square roots of probabilities, it suffices to repeat
the amplitude amplification process approximately $1/\sqrt{a_x}$ times to
achieve success with overwhelming probability.  For simplicity, we assume in
the rest of this paper that there is a single instance for which we seek a good
solution, which allows us to dispense with input~$x$, but the generalization to
the paradigm outlined above is straightforward.  Grover's original database
searching quantum algorithm is a special case of this process, in which $\chi$
is given by a function $f: \{0,1,\ldots, N-1\} \rightarrow \{0,1\}$
for which we are promised that there exists a unique $x_0$
such that \mbox{$f(x_0)=1$}.  If~we use the Fourier transform
as quantum algorithm~$\mathcal A$---or~more simply the Walsh--Hadamard
transform in case $N$ is a power of~2---an~equal superposition of all possible
$x$'s is produced, whose success probability would be~$1/N$
if measured.  Classical repetition would succeed after an expected number $N$
of evaluations of~$f$.  Amplitude amplification corresponds to Grover's
algorithm: it succeeds after approximately $\sqrt{N}$ evaluations of the
function.

We~generalize this result further to the case when the probability of success
$a$ of algorithm $\mathcal A$ is not known ahead of time: it remains sufficient
to evaluate $\mathcal A$ and $\chi$ an expected number of times that is
proportional to $1/\sqrt{a}$.  Moreover, in the case $a$ is known ahead of
time, we give two different techniques that are guaranteed to find a good
solution after a number of iterations that is proportional to $1/\sqrt{a}$
in the worst case.

It can be proven that Grover's algorithm goes quadratically faster than
any possible classical algorithm when function $f$ is given as a black~box.
However, it is usually the case in practice
that information is known about~$f$
that allows us to solve the problem much more efficiently than by exhaustive
search.  The~use of classical {\em heuristics}, in particular, will often
yield a solution significantly more efficiently than straight quantum
amplitude amplification would.  In~Section~\ref{sec:heuri}, we consider
a broad class of classical heuristics and show how to apply amplitude
amplification to obtain quadratic speedup compared to any such heuristic.

Finally, Section~\ref{sec:estimation} addresses the question of
estimating the success probability $a$ of quantum algorithm~$\mathcal A$.
We~call this process {\em amplitude estimation}.  As~a special case
of our main result (Theorem~\ref{amp_est}), an estimate for
$a$ is obtained after any number $M$ of iterations which is within
\mbox{$2\pi\sqrt{a(1-a)}/M+\pi^2/M^2$} of the correct value
with probability at least~$8/\pi^2$, where one iteration consists
of running algorithm $\mathcal A$ once forwards and once backwards,
and of computing function $\chi$ once.  As~an application of this
technique, we show how to approximately count the number of $x$
such that \mbox{$f(x)=1$} given a function
$f: \{0,1,\ldots, N-1\} \rightarrow \{0,1\}$.  If~the correct answer
is~\mbox{$t>0$}, it suffices to compute the function
$\sqrt{N}$ times to obtain an estimate roughly within $\sqrt{t}$
of the correct answer. A~number of evaluations of $f$ proportional
to $\oneovereps\sqrt{N/t}$ yields a result that is likely to
be within \mbox{$\eps t$} of the correct answer.
(We~can do slightly better in case $\eps$ is not fixed.)
If~it is known ahead of time that the correct answer is either
\mbox{$t=0$} or \mbox{$t=t_0$} for some fixed $t_0$, we can determine
which is the case with certainty using a number of evaluations of $f$
proportional to $\sqrt{N/t_0}$.  If~we have no prior knowledge about~$t$,
the exact count can
be obtained with high probability after a number of evaluations of $f$
that is proportional to~$\sqrt{t(N-t)}$
when \mbox{$0<t<N$} and $\sqrt{N}$ otherwise.
Most of these results are optimal.

We~assume in this paper that the reader is familiar with
basic notions of quantum computing.


\section{Quantum amplitude amplification}\label{sec:ampl}

Suppose we have a classical randomized algorithm that succeeds
with some probability~$p$.  If~we repeat the algorithm, say,
$j$~times, then our probability of success increases to
roughly~$jp$ (assuming $jp \ll 1$).
Intuitively, we can think of this strategy as each additional run
of the given algorithm boosting the probability of success by
an additive amount of roughly~$p$.

A~quantum analogue of boosting the probability of success would be
to boost the {\em amplitude\/} of being in a certain subspace
of a Hilbert space.  The general concept of amplifying the amplitude
of a subspace was discovered
by Brassard and H{\o}yer~\cite{BH} as a generalization of
the boosting technique applied by Grover in his
original quantum searching paper~\cite{Grover1}.
Following~\cite{BH} and~\cite{BBHT}, we refer to their idea as
{\em amplitude amplification\/} and detail the ingredients below.

Let $\mathcal H$ denote the Hilbert space representing
the state space of a quantum system.
Every Boolean function $\chi : \integer \rightarrow \{0,1\}$
induces a partition of~$\mathcal H$ into a direct sum
of two subspaces, a good subspace and a bad subspace.
The {\em good subspace\/} is the subspace spanned by
the set of basis states $\ket{x} \in {\mathcal H}$
for which $\chi(x)=1$,
and the {\em bad subspace\/} is its orthogonal complement
in~$\mathcal H$.
We~say that the elements of the good subspace are {\em good}, and
that the elements of the bad subspace are {\em bad}.

Every pure state $\ket{\Upsilon}$ in~$\mathcal H$ has a unique
decomposition as $\ket{\Upsilon} = \ket{\goodUpsilon} +
\ket{\badUpsilon}$, where \ket{\goodUpsilon} denotes the projection
onto the good subspace, and \ket{\badUpsilon} denotes the projection
onto the bad subspace.  Let $a_\Upsilon =
\braket{\goodUpsilon}$ denote the probability that measuring
\ket{\Upsilon} produces a good state, and similarly, let
$b_\Upsilon = \braket{\badUpsilon}$. Since \ket{\goodUpsilon} and
\ket{\badUpsilon} are orthogonal, we have
\mbox{$a_\Upsilon+b_\Upsilon=1$}.

Let $\mathcal A$ be any quantum algorithm that acts on~$\mathcal H$
and uses no measurements.
Let $\ket{\Psi} = {\mathcal A} \ket{0}$ denote the
state obtained by applying $\mathcal A$ to the initial zero state.
The amplification process is realized by repeatedly
applying the following unitary operator~\cite{BH}
on the state~\ket{\Psi},
\begin{equation}\label{eq:defq}
{\mathbf Q} = {\mathbf Q}({\mathcal A},\chi)
  = - {\mathcal A} \smallspace {\mathbf S}_0 \smallspace
    {\mathcal A}^{-1} \smallspace {\mathbf S}_\chi.
\end{equation}
Here, the operator ${\mathbf S}_\chi$ conditionally changes the
sign of the amplitudes of the good states,
\begin{equation*}
\ket{x} \;\longmapsto\; \begin{cases}
- \ket{x} & \text{if $\chi(x)=1$}\\
\hphantom{-} \ket{x} & \text{if $\chi(x)=0$,}\end{cases}
\end{equation*}
while the operator ${\mathbf S}_0$
changes the sign of the amplitude if and only
if the state is the zero state~\ket{0}.
The operator $\mathbf Q$ is well-defined since
we assume that $\mathcal A$ uses no measurements
and, therefore, $\mathcal A$ has an inverse.

The usefulness of operator $\mathbf Q$ stems from its simple action
on the subspace~${\mathcal H}_{\Psi}$ spanned by the vectors
\ket{\goodPsi} and~\ket{\badPsi}.
\begin{lemma}\label{lm:Qaction}
We have that
\begin{align*}
{\mathbf Q} \ket{\goodPsi} &\,=\, \hpsr{(1-2a)}{2(1-a)}
\ket{\goodPsi} - \hpsr{2a}{(1-2a)} \ket{\badPsi} \\
{\mathbf Q} \ket{\badPsi} &\,=\, 2(1-a) \ket{\goodPsi} + (1-2a)
\ket{\badPsi},
\end{align*}
where $a = \braket{\goodPsi}$.
\end{lemma}
It~follows that the subspace ${\mathcal H}_{\Psi}$ is stable
under the action of~$\mathbf Q$,
a property that was first observed by
Brassard and H{\o}yer~\cite{BH} and
rediscovered by Grover~\cite{Grover3}.

Suppose~$0<a<1$.  Then ${\mathcal H}_\Psi$ is a subspace of dimension~2,
and otherwise ${\mathcal H}_\Psi$ has dimension~1.
The action of~$\mathbf Q$ on ${\mathcal H}_\Psi$
is also realized by the operator
\begin{equation}\label{eq:reflections}
{\mathbf U}_\Psi {\mathbf U}_{\badPsi},
\end{equation}
which is composed of 2 reflections. The first operator, ${\mathbf
U}_{\badPsi}
= \identity - \frac{2}{1-a} \ketbra{\badPsi}$,
implements a reflection through the ray spanned by the vector
$\ket{\badPsi}$, while the second operator ${\mathbf U}_\Psi =
\identity - 2 \ketbra{\Psi}$ implements a reflection through the
ray spanned by the vector~$\ket{\Psi}$.

Consider the orthogonal complement~${\mathcal H}_\Psi^\perp$
of ${\mathcal H}_\Psi$ in~$\mathcal H$.
Since the operator
${\mathcal A} \smallspace {\mathbf S}_0 \smallspace
{\mathcal A}^{-1}$
acts as the identity on ${\mathcal H}_\Psi^\perp$,
operator $\mathbf Q$ acts as $- {\mathbf S}_\chi$
on~${\mathcal H}_\Psi^\perp$.
Thus, ${\mathbf Q}^2$ acts as the identity
on~${\mathcal H}_\Psi^\perp$, and
every eigenvector of~$\mathbf Q$ in ${\mathcal H}_\Psi^\perp$
has eigenvalue $+1$ or~$-1$.
It~follows that to understand the action of~$\mathbf Q$
on an arbitrary initial vector \ket{\Upsilon} in $\mathcal H$,
it suffices to consider the
action of~$\mathbf Q$ on the projection of \ket{\Upsilon}
onto~${\mathcal H}_\Psi$.

Since operator $\mathbf Q$ is unitary,
the subspace ${\mathcal H}_{\Psi}$ has an orthonormal basis
consisting of two eigenvectors of~$\mathbf Q$,
\begin{equation}\label{eq:eigenvectors}
\ket{\Psi_{\pm}} = \frac{1}{\sqrt 2}
\left(\frac{1}{\sqrt a}\ket{\goodPsi}
\pm \frac{\imath}{\sqrt{1-a}}\ket{\badPsi}\right),
\end{equation}
provided \mbox{$0 < a < 1$},
where $\imath = \sqrt{-1}$ denotes the principal square root
\mbox{of~$-1$}.  The corresponding eigenvalues are
\begin{equation}\label{eq:eigenvalues}
\lambda_\pm = e^{\pm \imath 2 \mytheta},
\end{equation}
where the angle~$\mytheta$ is defined so that
\begin{equation}\label{def_a}
\sin^2(\mytheta) = a = \braket{\goodPsi}
\end{equation}
and $0 \leq \mytheta \leq \pi/2$.

We use operator~$\mathbf Q$ to boost the success probability~$a$
of the quantum algorithm~$\mathcal A$.
First, express $\ket{\Psi} = \mathcal A \ket{0}$
in the eigenvector basis,
\begin{equation}\label{Q_at_A}
{\mathcal A} \ket{0} = \ket{\Psi}
= \frac{- \imath}{\sqrt 2} \left(e^{\imath \mytheta} \ket{\Psi_{+}}
- e^{- \imath \mytheta} \ket{\Psi_{-}}\right).
\end{equation}
It~is now immediate that after $j$ applications of
operator~$\mathbf Q$, the state~is
\begin{align}
{\mathbf Q}^j \ket{\Psi} &= \frac{- \imath}{\sqrt 2}
\left(e^{(2j+1) \imath \mytheta} \ket{\Psi_{+}}
- e^{- (2j+1) \imath \mytheta} \ket{\Psi_{-}}\right)
\label{eq:qjeigenvalues}\\
&= \frac{1}{\sqrt a} \sin((2j+1)\mytheta) \,\ket{\goodPsi}
 + \frac{1}{\sqrt{1-a}} \cos((2j+1)\mytheta) \,\ket{\badPsi}.
\label{eq:qjcompbasis}
\end{align}
It~follows that if $0<a<1$ and if we
compute ${\mathbf Q}^m \ket{\Psi}$ for some integer \mbox{$m \geq 0$},
then a final measurement will produce a good state
with probability equal to $\sin^2((2m+1)\mytheta)$.

If~the initial success probability~$a$ is either~0 or~1, then the
subspace ${\mathcal H}_\Psi$ spanned by \ket{\goodPsi}
and~\ket{\badPsi} has dimension~1 only, but the conclusion remains
the same: If~we measure the system after $m$~rounds of amplitude
amplification, then the outcome is good with probability
$\sin^2((2m+1)\mytheta)$, where the angle~$\mytheta$ is defined so
that Equation~\ref{def_a} is satisfied and so that $0 \leq \mytheta
\leq \pi/2$.

Therefore, assuming $a>0$,
to obtain a high probability of success, we want to
choose integer~$m$ such that $\sin^2((2m+1)\mytheta)$ is close to~1.
Unfortunately, our ability to choose $m$ wisely depends on our
knowledge about~$\mytheta$, which itself depends on~$a$.
The two extreme cases are when we know the exact
value of~$a$, and when we have no prior knowledge about $a$
whatsoever.

Suppose the value of $a$ is known.
If \mbox{$a > 0$}, then by letting
$m = \lfloor \pi /4 \mytheta \rfloor$, we have
that $\sin^2((2m+1)\mytheta) \geq 1 -a$, as shown in~\cite{BBHT}.
The next theorem is immediate.

\begin{theorem}[Quadratic speedup]\label{thm:quad}
Let $\mathcal A$ be any quantum algorithm that uses no measurements,
and let $\chi : \integer \rightarrow \{0,1\}$ be any Boolean
function.  Let $a$ the initial success probability of~$\mathcal A$.
Suppose $a>0$, and set $m = \lfloor \pi / 4 \mytheta \rfloor$,
where $\mytheta$ is defined so that $\sin^2(\mytheta) = a$ and
$0 < \mytheta \leq \pi/2$.
Then, if we compute ${\mathbf Q}^m {\mathcal A} \,\ket{0}$
and measure the system, the outcome is good with probability
at least \mbox{$\max(1-a,a)$}.
\end{theorem}

Note that any implementation of algorithm
${\mathbf Q}^m {\mathcal A} \ket{0}$ requires that the value of~$a$
is known so that the value of~$m$ can be computed.
We~refer to Theorem~\ref{thm:quad} as a quadratic speedup,
or the square-root running-time result.  The reason for this is that
if an algorithm $\mathcal A$ has success probability \mbox{$a>0$},
then after an expected number of $1/a$ applications of~$\mathcal A$,
we will find a good solution.
Applying the above theorem reduces this to an expected number of
at most $(2m+1)/\hspace{-.55mm}\max(1-a,a) \in \Theta(\sqrta)$
applications of $\mathcal A$ and~${\mathcal A}^{-1}$.

As~an application of Theorem~\ref{thm:quad}, consider the
search problem~\cite{Grover2} in which we are given
a Boolean function $f: \{0,1,\ldots,N-1\} \rightarrow \{0,1\}$
satisfying the promise that there exists
a unique $x_0 \in \{0,1,\ldots,N-1\}$
on which $f$ takes value~1, and we are asked to find~$x_0$.
If~$f$ is given as a black box, then on a classical computer,
we need to evaluate $f$ on an expected number of roughly half
the elements of the domain in order to determine~$x_0$.

By~contrast, Grover~\cite{Grover2} discovered
a quantum algorithm that only requires an expected number
of evaluations of $f$ in the order of~$\sqrt{N}$.
In~terms of amplitude amplification, Grover's algorithm reads
as follows:  Let $\chi = f$, and
let ${\mathcal A} = {\mathbf W}$ be the Walsh-Hadamard
transform on $n$ qubits that maps the initial zero state \ket{0}
to $\frac{1}{\sqrt{N}} \sum_{x=0}^{N-1} \ket{x}$,
an equally-weighted superposition of
all $N=2^n$ elements in the domain of~$f$.
Then the operator
${\mathbf Q} = - {\mathcal A} {\mathbf S}_0 {\mathcal A}^{-1}
{\mathbf S}_\chi$ is equal to the iterate
$- {\mathbf W} {\mathbf S}_0 {\mathbf W} {\mathbf S}_f$
applied by Grover in his searching paper~\cite{Grover2}.
The initial success probability $a$ of $\mathcal A$
is exactly $1/N$, and
if we measure after $m = \lfloor \pi /4 \mytheta\rfloor$
iterations of~$\mathbf Q$, the probability of measuring $x_0$
is lower bounded by \mbox{$1-1/N$}~\cite{BBHT}.

Now, suppose that the value of~$a$ is not known.
In~Section~\ref{sec:estimation}, we discuss techniques for finding
an estimate of~$a$, whereafter one then can apply a
weakened version of Theorem~\ref{thm:quad}
in which the exact value of~$a$ is replaced by an estimate of~it.
Another idea is to try to find a good solution without prior
computation of an estimate of~$a$.  Within that approach,
by adapting the ideas in Section~6 in~\cite{BBHT}
we can still obtain a quadratic speedup.

\begin{theorem}[Quadratic speedup without knowing~{\boldmath $a$}]
\label{thm:withouta}
There exists a quantum algorithm \algqsearch with the following property.
Let $\mathcal A$ be any quantum algorithm that uses no measurements,
and let $\chi : \integer \rightarrow \{0,1\}$ be any Boolean function.
Let $a$ denote the initial success probability of~$\mathcal A$.
Algorithm \algqsearch finds a good solution using an expected number
of applications of $\mathcal A$ and ${\mathcal A}^{-1}$ which are in
$\Theta(\sqrta)$ if $a>0$, and otherwise runs forever.
\end{theorem}

The algorithm in the above theorem utilizes the given quantum
algorithm~$\mathcal A$ as a subroutine and the operator~$\mathbf Q$.
The complete algorithm is as follows:

\begin{algorithm}{$\algqsearch(\mathcal{A},\chi)$}
\begin{enumerate}
\item  Set $l=0$ and let $c$ be any constant such that $1 < c <2$.
\item  Increase $l$ by~1 and set
       $M = \rblc{\upceil} c^l\rbrc{\upceil}$.
       \label{item:counterQsearch}
\item  Apply $\mathcal A$ on the initial state~\ket{0}, and
       measure the system.  If~the outcome \ket{z} is good, that is,
       if $\chi(z)=1$, then output~$z$ and stop.
       \label{item:technicality}
\item  Initialize a register of appropriate size to the state
       ${\mathcal A} \ket{0}$.
\item  Pick an integer $j$ between 1 and~$M$ uniformly at random.
\item  Apply ${\mathbf Q}^j$ to the register,
       where ${\mathbf Q} = {\mathbf Q}({\mathcal A},\chi)$.
\item  Measure the register.
       If~the outcome \ket{z} is good, then output~$z$ and stop.
       Otherwise, go~to step~\ref{item:counterQsearch}.
       \label{item:meas}
\end{enumerate}
\end{algorithm}

The intuition behind this algorithm is as follows.
In~a 2-dimensional real vector space, if we pick a
unit vector $(x,y) = (\cos(\cdot),\sin(\cdot))$
uniformly at random then the expected value of $y^2$ is~$1/2$.
Consider~Equation~\ref{eq:qjcompbasis}.
If we pick $j$ at random between 1 and~$M$ for some
integer~$M$ such that $M \theta_a$ is larger than, say, $100 \pi$,
then we have a good approximation to a random unit vector,
and we will succeed with probability close to~$1/2$.

To turn this intuition into an algorithm,
the only obstacle left is that we do not know the value of $\theta_a$,
and hence do not know an appropriate value for~$M$.
However, we can overcome this by using exponentially increasing
values of~$M$,
an idea similar to the one used in ``exponential searching''
(which is a~term that does not refer to the running time of the
method, but rather to an exponentially increasing growth of the size
of the search space).

The correctness of algorithm \algqsearch is immediate
and thus to prove
the theorem, it suffices to show that
the expected number of applications of~$\mathcal A$
and~${\mathcal A}^{-1}$ is in the order of~$1/\sqrt{a}$.
This can be proven by essentially the same techniques applied
in the proof of Theorem~3 in~\cite{BBHT} and we therefore
only give a very brief sketch of the proof.

On~the one hand, if~the initial success probability~$a$ is
at least~$3/4$, then step~\ref{item:technicality} ensures that
we soon will measure a good solution.
On~the other hand, if $0 < a < 3/4$ then,
for any given value of~$M$, the probability of measuring a
good solution in step~\ref{item:meas} is lower bounded by
\begin{equation}
\frac{1}{2}\left(1 - \frac{1}{2M\sqrt a}\right).
\end{equation}

Let $c_0>0$ be such that $c=2(1-c_0)$ 
and let $M_0 = 1/(2c_0 \sqrt{a}\,)$. 
The expected number of applications of~$\mathcal A$ is upper
bounded by $T_1 + T_2$, where 
$T_1$ denotes the maximum number of applications of~$\mathcal A$ 
the algorithm uses while $M < M_0$, 
and where
$T_2$ denotes the expected number of applications of~$\mathcal A$ 
the algorithm uses while $M \geq M_0$.
Clearly \mbox{$T_1 \in O(M_0) = O(\sqrta)$} 
and we now show that \smash{\mbox{$T_2 \in O(\sqrta)$}} as~well.
 
For all $M \geq M_0$, the measurement in step~\ref{item:meas} 
yields a good solution with probability at least~$\frac{1}{2}(1-c_0)$,
and hence it fails to yield a good solution with probability 
at most~$p_0 = \frac{1}{2}(1+c_0)$.
Thus for all $i\geq 0$,
with probability at most~$p_0^i$, we have that $M \geq M_0 c^i$ 
at some point after step~\ref{item:counterQsearch} 
while running the algorithm.
Hence $T_2$ is 
at most on the order of $\sum_{i \geq 0} M_0 (c p_0)^i$ 
which is in $O(M_0)$ since \mbox{$c p_0<1$}.
The total expected number of applications of~$\mathcal A$ 
is thus in~$O(M_0)$, which is $O(\sqrta)$.

For the lower bound, if $M$ were in $o\big(\sqrta\big)$,
then the probability
that we measure a good solution in step~\ref{item:meas} would be
vanishingly small.  
This completes our sketch of the proof of Theorem~\ref{thm:withouta}.

\subsection{Quantum de-randomization when the success\\probability is known}
\label{sec:knowna}
We~now consider the situation where the success probability~$a$
of the quantum algorithm~$\mathcal A$ is known.
If~$a=0$ or $a=1$,
then amplitude amplification will not change the success probability,
so in the rest of this section, we assume that \mbox{$0<a<1$}.
Theorem~\ref{thm:quad} allows us to boost the probability of
success to at least $\max(1-a,a)$.
A~natural question to ask is whether it is possible to improve this
to certainty, still given the value of~$a$.
It~turns out that the answer is positive.
This is unlike classical computers,
where no such general de-randomization technique is known.
We~now describe 2~optimal methods for obtaining this,
but other approaches are possible.

The first method is by applying amplitude amplification, not on the
original algorithm~$\mathcal A$, but on a slightly modified version
of it.  By Equation~\ref{eq:qjcompbasis},
if we measure the state ${\mathbf Q}^m {\mathcal A} \ket{0}$, then
the outcome is good with probability $\sin^2((2m+1)\mytheta)$.
In~particular,
if ${\tilde m} = \pi/4 \mytheta - 1/2$ happens to be an integer,
then we would succeed with certainty after $\tilde m$ applications
of~$\mathbf Q$.  In~general, ${\overline m}
 = \rblc{\upceil}{\tilde m}\rbrc{\upceil}$
iterations is a fraction of 1~iteration too many,
but we can compensate for that by choosing
$\barmytheta = \pi/(4 \overline m +2)$,
an angle slightly smaller than~$\mytheta$.  Any quantum
algorithm that succeeds with probability~$\overline a$ such
that \mbox{$\sin^2(\tinyspace\barmytheta\tinyspace) = \overline a$},
will succeed with certainty after $\overline m$ iterations of
amplitude amplification.
Given $\mathcal A$ and its initial success probability~$a$,
it is easy to construct a new quantum algorithm that succeeds
with probability \mbox{$\overline a \leq a$}:
Let $\mathcal B$ denote the quantum algorithm that takes a single
qubit in the initial state \ket{0} and
rotates it to the superposition
$\sqrt{1-\overline a/a\;\!}\;\! \ket{0}
 + \sqrt{\overline a/a\;\!}\;\! \ket{1}$.
Apply both $\mathcal A$ and $\mathcal B$, and define
a good solution as one in which $\mathcal A$ produces a good
solution, and the outcome of $\mathcal B$ is the state~\ket{1}.
Theorem~\ref{thm:witha} follows.

\begin{theorem}[Quadratic speedup with known~{\boldmath $a$}]
\label{thm:witha}
Let $\mathcal A$ be any quantum algorithm that uses no measurements,
and let $\chi : \integer \rightarrow \{0,1\}$ be any Boolean function.
There exists a quantum algorithm
that given the initial success probability~$a>0$ of $\mathcal A$,
finds a good solution with certainty
using a number of applications
of $\mathcal A$ and~${\mathcal A}^{-1}$ which is in $\Theta(\sqrta)$
in the worst case.
\end{theorem}

The second method to obtain success probability~1 requires
a generalization of operator~$\mathbf Q$.
Given angles $0 \leq \phi, \varphi < 2 \pi$,
redefine $\mathbf Q$ as follows,
\begin{equation}\label{eq:redefq}
{\mathbf Q} = {\mathbf Q}({\mathcal A},\chi,\phi,\varphi)
  = - {\mathcal A} \smallspace {\mathbf S}_0(\phi) \smallspace
    {\mathcal A}^{-1} \smallspace {\mathbf S}_\chi(\varphi).
\end{equation}
Here, the operator ${\mathbf S}_\chi(\varphi)$
is the natural generalization of the ${\mathbf S}_\chi$ operator,
\begin{equation*}
\ket{x} \;\longmapsto\; \begin{cases}
e^{\imath \varphi} \ket{x} & \text{if $\chi(x)=1$}\\
\hphantom{e^{\imath \varphi}} \ket{x}
  & \text{if $\chi(x)=0$.}\end{cases}
\end{equation*}
Similarly, the operator ${\mathbf S}_0(\phi)$
multiplies the amplitude by a factor of~$e^{\imath \phi}$
if and only if the state is the zero state~\ket{0}.
The action of operator ${\mathbf Q}({\mathcal A},\chi,\phi,\varphi)$
is also realized by applying an operator that is composed of
two pseudo-reflections: the operator
${\mathcal A} \smallspace {\mathbf S}_0(\phi) \smallspace
    {\mathcal A}^{-1}$
and the operator $-{\mathbf S}_\chi(\varphi)$.

The next lemma shows that the subspace ${\mathcal H}_\Psi$ spanned
by \ket{\goodPsi} and~\ket{\badPsi} is stable under the action
of~$\mathbf Q$, just as in the special case ${\mathbf Q}({\mathcal
A},\chi,\pi,\pi)$ studied above.

\begin{lemma}\label{lm:reQaction}
Let ${\mathbf Q} = {\mathbf Q}({\mathcal A},\chi,\phi,\varphi)$.
Then
\begin{align*}
{\mathbf Q} \ket{\goodPsi}
 &\,=\, e^{\imath \varphi}((1-{e^{\imath \phi}})a-1) \ket{\goodPsi}
     + \hpsr{e^{\imath \varphi}(1-{e^{\imath \phi}})
    a}{((1-{e^{\imath \phi}})a + {e^{\imath \phi}})} \ket{\badPsi}\\
{\mathbf Q} \ket{\badPsi}
 &\,=\, \hpsr{(1-{e^{\imath \phi}})(1-a)}{e^{\imath \varphi}((1-
    {e^{\imath \phi}})a-1)} \ket{\goodPsi}
     -((1-{e^{\imath \phi}})a + {e^{\imath \phi}}) \ket{\badPsi},
\end{align*}
where $a = \braket{\goodPsi}$.
\end{lemma}

Let $\tilde m = \pi/4 \mytheta - 1/2$, and suppose that $\tilde m$ is
not an integer.  In~the second method to obtain a good solution
with certainty, we also apply
$\rblc{\upceil}{\tilde m}\rbrc{\upceil}$
iterations of amplitude amplification, but now we slow down the speed
of the very last iteration only, as opposed to of all iterations as in
the first method.  For the case $\tilde m < 1$, this second method
has also been suggested by Chi and Kim~\cite{CK}.
We~start by applying the
operator~$\mathbf Q({\mathcal A},\chi,\phi,\varphi)$
with $\phi = \varphi = \pi$
a~number of $\lfloor \tilde m\rfloor$ times to
the initial state $\ket{\Psi} = {\mathcal A} \ket{0}$.
By~Equation~\ref{eq:qjcompbasis}, this produces the superposition
\begin{equation*}
\frac{1}{\sqrt a} \sin\big((2 \lfloor \tilde m \rfloor+1)\mytheta\big)
\,\ket{\goodPsi}
+ \frac{1}{\sqrt{1-a}} \cos\big((2\lfloor \tilde m\rfloor+1)\mytheta\big)
\,\ket{\badPsi}.
\end{equation*}
Then, we apply operator~$\mathbf Q$ one more time,
but now using angles $\phi$ and $\varphi$, both between $0$ and~$2\pi$,
satisfying
\begin{multline}\label{eq:lastmeal}
e^{\imath \varphi} (1-e^{\imath \phi}){\sqrt a}
\sin\big((2 \lfloor \tilde m \rfloor +1) \mytheta\big)\\
  = ((1-e^{\imath \phi})a + e^{\imath \phi}) \frac{1}{\sqrt{1-a}}
    \cos\big((2 \lfloor \tilde m \rfloor +1) \mytheta\big) \,.
\end{multline}
By~Lemma~\ref{lm:reQaction}, this ensures that the resulting
superposition has inner product zero with \ket{\badPsi}, and thus a
subsequent measurement will yield a good solution with certainty.

The problem of choosing $\phi,\varphi \in \reals$
such that
Equation~\ref{eq:lastmeal} holds
is equivalent to requiring that
\begin{equation}\label{eq:lastmeal2}
\cot\big((2 \lfloor \tilde m \rfloor +1) \mytheta\big)
= e^{\imath \varphi} \sin(2 \mytheta)
  \big( -\cos(2 \mytheta)
  + \imath \cot(\phi/2)\big)^{-1}.
\end{equation}
By~appropriate choices of $\phi$ and~$\varphi$, the right
hand side of Equation~\ref{eq:lastmeal2} can be made equal to
any nonzero complex number of norm at most~$\tan(2 \mytheta)$.
Thus, since the left hand side of this equation is equal to
some real number smaller than~$\tan(2 \mytheta)$,
there exist $\phi, \varphi \in \reals$
such that Equation~\ref{eq:lastmeal2} is satisfied,
and hence also such that
the expression in Equation~\ref{eq:lastmeal} vanishes.
In~conclusion, applying ${\mathbf Q}({\mathcal A},\chi,\phi,\varphi)$
with such $\phi, \varphi \in \reals$ at the very last iteration
allows us to measure a good solution with certainty.


\section{Heuristics}\label{sec:heuri}

As explained in the previous section, using
the amplitude amplification
technique to search for a solution to a search problem,
one obtains a quadratic speedup compared to
a brute force search.
For many problems, however, good heuristics are known for which
the expected running time, when applied to a ``real-life'' problem,
is in $o(\sqrt{N})$,
where $N$ is the size of the search space.
This fact would make amplitude amplification much less useful
unless a quantum computer is somehow able to take advantage
of these classical heuristics.
In this section we concentrate on a large family of
classical heuristics that can be applied to search problems.
We show how these heuristics can be incorporated into the
general amplitude amplification process.

By a heuristic, we mean a probabilistic algorithm, running
in polynomial time, that outputs what one is searching for with
some non-negligible probability.

Suppose we have a family~${\mathcal F}$ of functions
such that each \mbox{$f \in {\mathcal F}$} is of the form
$f:X \rightarrow \{0,1\}$.  For a given function $f$
we seek an input $x \in X$ such that $f(x)=1$.
A~{\em heuristic\/} is a function
$G: {\mathcal F} \times R \rightarrow X$, for an appropriate
finite set~$R$.  The heuristic $G$ uses a random seed
$r\in R$ to generate a guess for an $x$ such that $f(x)=1$.
For every function $f \in {\mathcal F}$, let
$t_f=|\{x \in X \mid f(x)=1\}|$, the number of good inputs $x$,
and let
\mbox{$h_f=|\{r \in R \mid f(G(f,r))=1\}|$}, the number
of good seeds.
We say that the heuristic is {\em efficient\/} for a given $f$ if
$h_f/|R|  > t_f/|X|$, that is, if using $G$ and a random
seed to generate inputs to $f$ succeeds with a higher
probability than directly guessing inputs to $f$ uniformly at random.
The heuristic is {\em good\/} in general if
\[\textup{E}_{\mathcal{F}}\left( \frac{h_f}{|R|} \right) \ > \
\textup{E}_{\mathcal{F}}\left( \frac{t_f}{|X|} \right) \ .\]
Here $\textup{E}_{\mathcal{F}}$ denotes the expectation over all $f$ according
to some fixed distribution.
Note that for some $f$, $h_f$ might be small but
repeated uses of the heuristic,
with seeds uniformly chosen in~$R$,
will increase the probability of finding a solution.

\begin{theorem}\label{thm:heu}
Let ${\mathcal F} \subseteq \{f \mid f:X \rightarrow \{0,1\}\}$
be a family of
Boolean functions and $\mathcal D$ be a probability distribution
over ${\mathcal F}$.
If on a classical computer,
using heuristic $G:{\mathcal F} \times R \rightarrow X$,
one finds $x_0 \in X$ such that $f(x_0)=1$
for random $f$ taken from distribution $D$ in expected time $T$ then
using a quantum computer,
a solution can be found in expected time
in~$O(\sqrt{T}\,)$.
\end{theorem}

\begin{proof}
A simple solution to this problem is to embed
the classical heuristic $G$ into the function
used in the algorithm \algqsearch.
Let $\chi(r) = f(G(f,r))$ and
$x = G(f,\algqsearch({\mathbf W},\chi))$,
so that $f(x)=1$.
By~Theorem~\ref{thm:withouta},
for each function
$f \in {\mathcal F}$, we have an
expected running time in $\Theta(\sqrt{|R|/h_f}\,)$.
Let $P_f$ denote the probability that $f$ occurs.
Then \mbox{$\sum_{f \in \mathcal{F}} P_f =1$}, and
we have that the expected running time is
in the order of
$\sum_{f \in {\mathcal F}} \sqrt{|R|/h_f\;\!}\;\! P_f$, which can be
rewritten as
\[\sum_{f \in {\mathcal F}}   \sqrt{\frac{|R|}{h_f} P_f} \sqrt{P_f}
 \leq    \left(\sum_{f\in {\mathcal F}} \frac{|R|}{h_f} P_f \right)^{1/2}
\left(\sum_{f \in {\mathcal F}} P_f \right)^{1/2}
  =   \left(\sum_{f \in {\mathcal F}} \frac{|R|}{h_f} P_f \right)^{1/2} \]
by Cauchy--Schwarz's inequality.
\end{proof}

An alternative way to prove Theorem~\ref{thm:heu} is
to incorporate the heuristic into the operator
$\mathcal A$ and do a minor modification to~$f$.
Let $\mathcal{A}$ be the quantum implementation of~$G$.
It is required that the operator $\mathcal{A}$
be unitary, but clearly in general the classical heuristic
does not need to be reversible.
As~usual in quantum algorithms one will need first to
modify the heuristic \mbox{$G:{\mathcal F} \times R \rightarrow X$}
to make it reversible, which can be done efficiently
using standard techniques~\cite{Bennett}.
We obtain a reversible function
$G'_f: R \times \mathbf{0} \rightarrow R \times X$.
Let $\mathcal A$ be the natural unitary operation implementing
$G'_f$ and let us modify ${\chi}$ (the good set membership function)
to consider only the second part of the register,
that is $\chi( (r,x) )=1$ if and only if $f(x)=1$.
We then have that $a=h_f/|R|$ and by~Theorem~\ref{thm:withouta},
for each function
$f \in {\mathcal F}$, we have an
expected running time in $\Theta(\sqrt{|R|/h_f}\,)$.
The rest of the reasoning is similar.
This alternative technique shows, using a simple
example, the usefulness of the
general scheme of amplitude amplification
described in the preceding section, although it is
clear that from a computational point of view
this is strictly equivalent to the technique
given in the earlier proof of the theorem.


\section{Quantum amplitude estimation}\label{sec:estimation}

Section~\ref{sec:ampl} dealt in a very general way
with combinatorial search problems, namely,
given a Boolean function $f: X \rightarrow \{0,1\}$
find an $x \in X$ such that $f(x)=1$.
In~this section, we deal with the
related problem of estimating $t = | \{x \in X \mid f(x)=1\}|$,
the number of inputs on which $f$ takes the value~1.

We can describe this counting problem in terms of amplitude
estimation. Using the notation of Section~\ref{sec:ampl}, given a
unitary transformation ${\mathcal A}$ and a Boolean
function~$\chi$, let $\ket{\Psi} = {\mathcal A} \ket{0}$. Write
$\ket{\Psi} = \ket{\goodPsi} + \ket{\badPsi}$ as a superposition of
the good and bad components of~$\ket{\Psi}$. Then {\em amplitude
estimation\/} is the problem of estimating~$a = \braket{\goodPsi}$,
the probability that a measurement of~$\ket{\Psi}$ yields a good
state.

The problem of estimating $t = |\{x \in X \mid f(x)=1\}|$ can be
formulated in these terms as follows.
For simplicity, we take $X=\{0, 1,\ldots, N-1\}$.
If $N$ is a power of~2, then we set $\chi = f$ and
${\mathcal A}={\mathbf W}$.
If~$N$ is not a power of~2, we set $\chi =f$ and
${\mathcal A}={\mathbf F}_N$,
the quantum Fourier transform which,
for every integer $M \geq 1$, is defined by
\begin{equation}\label{eq:cyclicFT}
{\mathbf F}_M \;:\; \ket{x} \;\longmapsto\;
  \frac{1}{\sqrt M} \sum_{y=0}^{M-1}
  e^{2 \pi \imath x y/M} \ket{y} \qquad (0 \leq x < M).
\end{equation}
Then in both cases we have $a = t/N$,
and thus an estimate for $a$ directly translates into an estimate
for~$t$.

To~estimate~$a$, we make good use of the properties of operator
${\mathbf Q} =
- {\mathcal A}\smallspace{\mathbf S}_0\smallspace{\mathcal A}^{-1}
\smallspace{\mathbf S}_f$.
By Equation~\ref{eq:qjcompbasis} in Section~\ref{sec:ampl}, we have
that the amplitudes of $\ket{\goodPsi}$ and $\ket{\badPsi}$ as
functions of the number of applications of~${\mathbf Q}$, are
sinusoidal functions, both of period~$\frac{\pi}{\theta_a}$. Recall
that $0 \leq \theta_a \leq \pi/2$ and $a = \sin^2(\theta_a)$, and
thus an estimate for $\theta_a$ also gives an estimate for~$a$.

To estimate this period, it is a natural approach~\cite{BHT} to
apply Fourier analysis like Shor~\cite{Shor} does for a classical
function in his factoring algorithm. This approach can also
be viewed as an eigenvalue estimation~\cite{Kitaev, CEMM} and is
best analysed in the basis of eigenvectors of the operator at hand~\cite{Mosca}.
By Equation~\ref{eq:eigenvalues}, the eigenvalues of ${\mathbf Q}$
on the subspace spanned by $\ket{\goodPsi}$ and $\ket{\badPsi}$ are
$\lambda_+ = e^{\imath 2 \theta_a}$ and $\lambda_- =e^{- \imath 2
\theta_a}$. Thus we can estimate $a$ simply by estimating one of
these two eigenvalues. Errors in our estimate $\tildetheta_a$ for
$\theta_a$ translate into errors in our estimate $\tilde{a} =
\sin^2(\tildetheta_a)$ for~$a$, as described in the next lemma.

\begin{lemma} \label{phase_to_amp}
Let $a=\sin^2(\theta_a)$ and $\tilde{a}=\sin^2(\tildetheta_a)$ with
$0 \leq \theta_a , \tildetheta_a \leq 2 \pi$ then
\begin{equation*}
\big| \tildetheta_a  - {\theta_a} \big| \leq \eps
{\ } \Rightarrow {\ }
   | \tilde{a} - a |
  \leq 2 \eps \sqrt{a(1-a)} + \eps^2\,.
\end{equation*}
\end{lemma}

\begin{proof}
For $\eps \geq 0$, using standard trigonometric identities, we obtain
\begin{eqnarray*}
\sin^2(\theta_a + \eps) - \sin^2(\theta_a)
&=&
\sqrt{a(1-a)}\sin(2\eps) + (1-2a) \sin^2(\eps) \textrm{ and}\\
\sin^2(\theta_a) - \sin^2(\theta_a - \eps) &=&
\sqrt{a(1-a)}\sin(2\eps) + (2a-1) \sin^2(\eps).
\end{eqnarray*}
The inequality follows directly.
\end{proof}

We~want to estimate one of the eigenvalues of~${\mathbf Q}$.
For this purpose, we utilize the following operator~$\Lambda$.
For any positive integer~$M$ and any unitary operator~${\mathbf{U}}$,
the operator $\Lambda_M({\mathbf U})$ is defined by
\begin{equation} \label{eqlambda}
\ket{j}\ket{y} \;\longmapsto\;
\ket{j}({\mathbf U}^j\ket{y}) \qquad (0 \leq j < M).
\end{equation}
Note that if $\ket{\Phi}$ is an eigenvector of ${\mathbf U}$ with eigenvalue
$e^{2 \pi \imath \omega}$, then
$\Lambda_M({\mathbf U})$ maps $\ket{j}\ket{\Phi}$
to $e^{2 \pi \imath \omega j}\ket{j}\ket{\Phi}$.

\begin{definition}
For any integer $M > 0$ and real number $0 \leq \omega < 1$, let
\[ \KS{\omega}{M} = \frac{1}{\sqrt{M}}
\sum_{y=0}^{M-1} e^{2 \pi \imath \omega  y} \,\ket{y}
.\]
We then have, for all $0 \leq x \leq M-1$
\[ \mathbf{F}_M \,\ket{x} = \KS{x/M}{M}.  \]
\end{definition}

The state
$\KS{\omega}{M}$ encodes the angle $2 \pi \omega$
($0 \leq \omega < 1$)
in the phases of
an equally weighted superposition of all basis states.
Different angles have different encodings, and the overlap between
$\KS{\omega_0}{M}$ and $\KS{\omega_1}{M}$
is a measure for the distance between the two
angles $\omega_0$ and~$\omega_1$.

\begin{definition}
For any two real numbers $\omega_0, \omega_1 \in \reals$, let
$d(\omega_0,\omega_1)
  = \min_{z \in \integer}\{|z+\omega_1-\omega_0| \}$.
\end{definition}

Thus $2\pi d(\omega_0,\omega_1)$ is the length of the shortest arc
on the unit circle going from $e^{2\pi \imath \omega_0}$
to~$e^{2 \pi \imath \omega_1}$.

\begin{lemma} \label{SS}
For $0 \leq \omega_0 < 1$ and $0 \leq \omega_1 < 1$ let
$\Delta = d(\omega_0, \omega_1)$.
If~$\Delta=0$  we have
$\left| \BKS{\omega_0}{\omega_1}{M}  \right|^2 = 1$.
Otherwise
\[ \left|  \BKS{\omega_0}{\omega_1}{M}  \right|^2 =
\frac{\sin^2( M \Delta \pi  ) }{M^2 \sin^2(\Delta \pi)}.  \]
\end{lemma}

\begin{proof}
\begin{eqnarray*}
\left|  \BKS{\omega_0}{\omega_1}{M} \right|^2
& = & \left| \left( \frac{1}{\sqrt{M}} \sum_{y=0}^{M-1}
e^{-2 \pi \imath \omega_0 y}
\bra{y}\right)\left(
\frac{1}{\sqrt{M}} \sum_{y=0}^{M-1} e^{2 \pi \imath \omega_1 y}
\ket{y}   \right) \right|^2 \\
& = &\frac{1}{M^2} \left|  \sum_{y=0}^{M-1} e^{2 \pi \imath \Delta y}
 \right|^2  \ = \ \frac{\sin^2( M \Delta \pi  ) }{M^2 \sin^2(\Delta \pi)}.\\
\end{eqnarray*}
\end{proof}

Consider the problem of
estimating $\omega$ where $0 \leq \omega <1$,
given the state $\KS{\omega}{M}$.
If $\omega = x/M$ for some integer $0 \leq x < M$,
then $\mathbf{F}^{-1}_M \KS{x/M}{M} = \ket{x}$ by definition,
and thus we have a perfect phase estimator.
If $M \omega$ is not an integer, then
observing $\mathbf{F}^{-1}_M \KS{\omega}{M}$ still provides a good
estimation of~$\omega$, as shown in the following theorem.

\begin{theorem} \label{QFT}
Let $X$ be the discrete random variable corresponding to the classical
result of measuring $\mathbf{F}^{-1}_M
\KS{\omega}{M}$ in the computational basis.  If $M \omega$  is an
integer then $ {\rm Prob}(X  = M \omega )=1 $.  Otherwise, letting
$\Delta = d(\omega,x/M)$,
\[ {\rm Prob}(X = x) \ = \ \frac{\sin^2( M
\Delta \pi  ) }{M^2 \sin^2(\Delta \pi)} \ \leq \  \frac{1}{(2 M \Delta)^2}\,.\]
For any $k>1$ we also have
\[ {\rm Prob}\left( d(X/M, \omega )  \leq  k/M \right)
 \  \geq \  1- \frac{1}{2 (k-1)} \]
and, in the case $k=1$ and $M > 2$,
\[ {\rm Prob}\left( d(X/M, \omega)  \leq  1/M \right)
 \  \geq \  \frac{8}{\pi^2} \, .  \]
\end{theorem}

\begin{proof}
Clearly
\begin{align*}
{\rm Prob}(X=x)
&\;=\; \left|  \bra{x}
     {\mathbf F}^{-1} \KS{\omega}{M} \right|^2\\
&\;=\; \left|  ({\mathbf F} \ket{x})^{\dagger} \KS{\omega}{M}  \right|^2\\
&\;=\; \left|  \BKS{x/M}{\omega}{M}  \right|^2
\end{align*}
thus using Lemma~\ref{SS} we directly obtain the first part of the
theorem.   We use this fact to prove the next part of the theorem.
\begin{eqnarray*}
{\rm Prob}\left( d(X/M,\omega)  \leq k/M
\right) &=& 1- {\rm Prob}( d(X/M,\omega) >  k/M ) \\ &\geq& 1- 2
\sum_{j=k}^{\infty} \frac{1}{4 M^2 (\frac{j}{M})^2 } \\ &\geq& 1-
\frac{1}{2 (k-1)}.
\end{eqnarray*}

For the last part, we use the fact that for $M>2$, the given expression
attains its minimum at $\Delta = 1/(2M)$
in the range \mbox{$0 \leq \Delta \leq 1/M$}.
\begin{eqnarray*}
{\rm Prob}\left( d(X/M,\omega) \leq
 1/M \right) &=& {\rm
Prob}(X = \lfloor M \omega \rfloor ) +
    {\rm Prob}(X = \lceil M \omega \rceil) \\
&=& \frac{\sin^2( M \Delta \pi  ) }{M^2 \sin^2(\Delta \pi)}+
\frac{\sin^2( M (\frac{1}{M} -\Delta) \pi  ) }
{M^2 \sin^2((\frac{1}{M} -\Delta) \pi)} \\
&\geq&
\frac{8}{\pi^2}.
\end{eqnarray*}
\end{proof}

The following algorithm computes an estimate for~$a$,
via an estimate for~$\theta_a$.

\begin{algorithm}{$\algestamp(\mathcal{A},\chi,M)$}
\begin{enumerate}
\item  Initialize two registers of appropriate sizes
to the state $\ket{0}{\mathcal{A}}\ket{0}$. \label{estInitstep}
\item\label{practice}  Apply $\mathbf{F}_M$ to the first register.
       \label{estFMstep}
\item  Apply $\Lambda_M({\mathbf Q})$ where ${\mathbf Q}=
- {\mathcal A}\smallspace{\mathbf S}_0\smallspace{\mathcal A}^{-1}
\smallspace{\mathbf S}_{\chi}$.
\item  Apply $\mathbf{F}_M^{-1}$ to the first register.
\item  Measure the first register and denote the outcome~$\ket{y}$.
    \label{lastitemtofigure}
\item  Output $ \tilde{a} = \sin^2 (\pi \frac{y}{M})$.
\end{enumerate}
\end{algorithm}

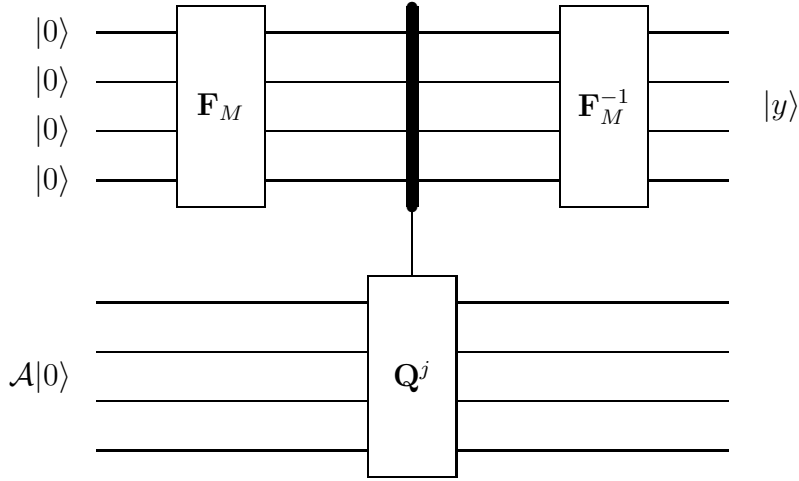
\begin{figure}[htbp]
\begin{center}
\setlength{\unitlength}{2160sp}
\begin{picture}(8000,6000)(1450,-5800)
\put(4900,-5750){\framebox(1000,2300)[c]{$\mathbf Q^j$}}%
\put(2700,-2650){\framebox(1000,2300)[c]{${\mathbf F}_M$}}%
\put(7100,-2650){\framebox(1000,2300)[c]{${\mathbf F}_M^{-1}$}}%
\put(1480,-2350){\makebox(0,0)[r]{$\ket{0}$}}
\put(1480,-1784){\makebox(0,0)[r]{$\ket{0}$}}
\put(1480,-1217){\makebox(0,0)[r]{$\ket{0}$}}
\put(1480,-650){\makebox(0,0)[r]{$\ket{0}$}}
\put(1480,-5750){\makebox(0,2300)[rc]{${\mathcal A} \ket{0}$}}
\put(9420,-2650){\makebox(0,2300)[lc]{$\ket{y}$}}
\put(2700,-2350){\line(-1,0){930}}
\put(2700,-1784){\line(-1,0){930}}
\put(2700,-1217){\line(-1,0){930}}
\put(2700,-650){\line(-1,0){930}}
\put(8100,-2350){\line(1,0){930}}
\put(8100,-1784){\line(1,0){930}}
\put(8100,-1217){\line(1,0){930}}
\put(8100,-650){\line(1,0){930}}
\put(3700,-2350){\line(1,0){3400}}
\put(3700,-1784){\line(1,0){3400}}
\put(3700,-1217){\line(1,0){3400}}
\put(3700,-650){\line(1,0){3400}}
\put(4900,-5450){\line(-1,0){3130}}
\put(4900,-4884){\line(-1,0){3130}}
\put(4900,-4317){\line(-1,0){3130}}
\put(4900,-3750){\line(-1,0){3130}}
\put(5900,-5450){\line(1,0){3130}}
\put(5900,-4884){\line(1,0){3130}}
\put(5900,-4317){\line(1,0){3130}}
\put(5900,-3750){\line(1,0){3130}}
\put(5400,-3450){\line(0,1){2800}}
\put(5461,-2650){\makebox(0,0)[c]{\circle*{136}}}
\put(5461,-350){\makebox(0,0)[c]{\circle*{136}}}
\put(5335,-2650){\makebox(0,0)[lb]{\rule{4.3pt}{75.805664pt}}}
\end{picture}
\caption{Quantum circuit for amplitude estimation.}
\label{fig:counting_fig}
\end{center}
\end{figure}

Steps~1 to~\ref{lastitemtofigure} are illustrated on
Figure~\ref{fig:counting_fig}.
This algorithm can also be summarized, following the
approach in~\cite{Hoyer}, as the unitary transformation
\[ \Big(({\mathbf F}_M^{-1}\otimes {\mathbf I}) \ \Lambda_M({\mathbf Q}) \
({\mathbf F}_M \otimes {\mathbf I})\Big)  \]
applied on state $\ket{0}{\mathcal A}\ket{0}$,
followed by a measurement of the first register and
classical post-processing of the outcome.
In~practice, we could choose $M$ to be a power of~2, which would allow us
to use a Walsh--Hadamard transform instead of a Fourier transform in
step~\ref{practice}.

\begin{theorem}[Amplitude Estimation] \label{amp_est}
For any positive integer $k$,
the algorithm $\algestamp({\mathcal A},\chi,M)$
outputs $\tilde{a}$ $(0 \leq \tilde{a} \leq 1)$ such that
\[ \left| \tilde{a} - a \right| \;\leq\;
2 \pi k \frac{\sqrt{a(1-a)}}{M} +  k^2 \frac{\pi^2}{M^2} \] with
probability at least $\frac{8}{\pi^2}$ when $k=1$ and
with probability greater than $1-\frac{1}{2(k-1)}$ for $k
\geq 2$.  It uses exactly $M$ evaluations of $f$.
If $a=0$ then ${\tilde a} =0$ with certainty, and
if $a=1$ and $M$ is even, then ${\tilde a} =1$ with certainty.
\end{theorem}

\begin{proof}
After step~\ref{estInitstep}, by Equation~\ref{Q_at_A}, we have state
\begin{eqnarray*}
\ket{0}{\mathcal{A}}\ket{0}  &=&
\frac{-\imath}{\sqrt 2} \ket{0}
\left(e^{\imath \mytheta}  \ket{\Psi_{+}}
 \, - \, e^{-\imath \mytheta}  \ket{\Psi_{-}}\right).
\end{eqnarray*}
 After step~\ref{estFMstep}, ignoring global phase, we have
\[
  \frac{1}{\sqrt{2M}}\sum_{j=0}^{M-1}\ket{j}  \left(e^{\imath
\mytheta} \ket{\Psi_{+}} \, - \, e^{-
\imath \mytheta} \ket{\Psi_{-}}\right)
\]
and after applying $\Lambda_M({\mathbf Q})$ we have
\begin{align*}
\hphantom{=}& \;
\frac{1}{\sqrt{2M}}\sum_{j=0}^{M-1}\ket{j}  \left(e^{\imath
\mytheta} e^{2 \imath j \mytheta}
\ket{\Psi_{+}} \, - \,  e^{- \imath \mytheta} e^{-2 \imath j \mytheta}
\ket{\Psi_{-}}\right) \\
=&\;
\frac{e^{\imath\mytheta}}{\sqrt{2M}}\sum_{j=0}^{M-1} e^{2
\imath j
\mytheta} \ket{j} \ket{\Psi_{+}}
{\ } - {\ }
 \frac{e^{-\imath\mytheta}}{\sqrt{2M}}\sum_{j=0}^{M-1}
e^{-2 \imath  j \mytheta} \ket{j}\ket{\Psi_{-}} \\
=&\;
\frac{e^{\imath\mytheta}}{\sqrt{2}}
 \KS{\frac{\theta_a}{\pi}}{M}\ket{\Psi_{+}} {\ } - {\ }
\frac{e^{-\imath\mytheta}}{\sqrt{2}}
 \KS{1-\frac{\theta_a}{\pi}}{M}\ket{\Psi_{-}}.
\end{align*}
We then apply ${\mathbf F}_M^{-1}$ to the first register and
measure it in the computational basis.

The rest of the proof follows from Theorem~\ref{QFT}. Tracing out
the second register in the eigenvector basis, we see that the first
register is in an equally weighted mixture of ${\mathbf
F}_M^{-1}\KS{\frac{\theta_a}{\pi}}{M}$ and ${\mathbf
F}_M^{-1}\KS{1-\frac{\theta_a}{\pi}}{M}$. Thus the measured value
$\ket{y}$ is the result of measuring either the state
  ${\mathbf F}_M^{-1}\KS{\frac{\theta_a}{\pi}}{M}$ or the state
${\mathbf F}_M^{-1} \KS{1-\frac{\theta_a}{\pi}}{M}$. The
probability of measuring $\ket{y}$ given the state ${\mathbf
F}_M^{-1}\KS{1-\frac{\theta_a}{\pi}}{M}$ is equal to the
probability of measuring $\ket{M - y}$ given the state ${\mathbf
F}_M^{-1}\KS{\frac{\theta_a}{\pi}}{M}$.  Since $\sin^2\big(\pi
\frac{(M-y)}{M}\big)
= \sin^2\big(\pi \frac{y}{M}\big)$, we can assume we measured
$\ket{y}$ given the state ${\mathbf F}_M^{-1}
\KS{\frac{\theta_a}{\pi}}{M}$ and $\tildetheta_a =
\pi \frac{y}{M}$ estimates $\theta_a$ as described in
Theorem~\ref{QFT}.
Thus we obtain bounds on $d(\tildetheta_a, \theta_a)$ that
translate, using Lemma~\ref{phase_to_amp}, into the appropriate
bounds on $|\tilde{a} - a|$.
\end{proof}

A~straightforward application of this algorithm is to approximately
count the number of solutions $t$ to $f(x)=1$.  To do this we
simply set ${\mathcal A} = {\mathbf W}$ if $N$ is a power of~2, or
in general ${\mathcal A} = {\mathbf F}_N$ or any other
transformation that maps $\ket{0}$ to
$\frac{1}{\sqrt{N}}\sum_{j=0}^{N-1} \ket{j}$. Setting $\chi=f$, we
then have $a = \braket{\goodPsi} = t/N$,
which suggests the following algorithm.

\begin{algorithm}{$\algcount(f,M)$} \label{count_alg}
\begin{enumerate}
\item  Output $t^{\prime} =N \ \times \
\algestamp({\mathbf F}_N,f,M)$.
\end{enumerate}
\end{algorithm}

By~Theorem~\ref{amp_est}, we obtain the following.

\begin{theorem}[Counting]\label{thm:count}
For any positive integers $M$ and~$k$,
and any Boolean function
$f:\{0,1,\ldots,N-1\} \rightarrow \{0,1\}$, the algorithm
\textup{\textbf{Count}}$(f,M)$ outputs an estimate $t^{\prime}$
to $t = |f^{-1}(1)|$ such that
\[ \big| t^{\prime} - t \big| \;\leq\;
2 \pi k \frac{\sqrt{t (N-t)}}{M} + \pi^2 k^2 \frac{N}{M^2} \]
with probability at least ${8}/{\pi^2}$ when $k=1$, and with
probability greater than $1-\frac{1}{2(k-1)}$ for $k \geq 2$. If
$t=0$ then $t^{\prime} =0$ with certainty, and if $t=N$ and $M$ is
even, then $t^{\prime} =N$ with certainty.
\end{theorem}

Note that \textup{\textbf{Count}}$(f,M)$ outputs a real number. In
the following counting algorithms we will wish to output an
integer, and therefore we will round off the output of
\textup{\textbf{Count}} to an integer.  To assure that the
rounding off can be done efficiently\footnote{For example, if
$t^{\prime} +
\frac{1}{2}$ is super-exponentially close to an integer $n$ we may not
be able to decide efficiently if $t^{\prime}$ is closer to $n$ or
$n-1$. } we will round off to an integer $\tilde{t}$ satisfying
$\left|
\tilde{t}
-
\mbox{\textup{\textbf{Count}}}(f,M) \right| \leq \frac{2}{3}$.

If we want to estimate $t$ within a few standard
deviations, we can apply algorithm \textbf{Count} with $M =
\raisebox{\upbceil}{$\big\lceil$} \sqrt{N} \,
   \raisebox{\upbceil}{$\big\rceil$}$.

\begin{corollary}\label{coro:std}
Given a Boolean function $f:\{0,1,\ldots,N-1\} \rightarrow \{0,1\}$
with $t$ defined as above, rounding off the output of
$\textup{\textbf{Count}}\big(f,\raisebox{\upbceil}{$\big\lceil$}
\sqrt{N} \,
   \raisebox{\upbceil}{$\big\rceil$}\,\big)$
gives an estimate~$\tilde{t}$ such that
\begin{equation}\label{eq:corostd}
\big|\tilde{t}-t\big| \;<\;
  2 \pi \sqrt{\frac{t(N-t)}{N}} + 11
\end{equation}
with probability at least
$8/\pi^2$ and requires
exactly $\raisebox{\upbceil}{$\big\lceil$} \sqrt{N} \,
   \raisebox{\upbceil}{$\big\rceil$}$
evaluations of~$f$.
\end{corollary}

We now look at the case of estimating $t$ with some relative error,
also referred to as {\em approximately counting $t$ with
accuracy~$\eps$}. For this we require the following crucial
observation about the output~$t^{\prime}$ of
algorithm $\textbf{Count}(f,L)$. Namely $t^{\prime}$
is likely to be equal to zero if and only if $L \in
o(\sqrt{N/t})$. Thus, we can find a rough estimate of $\sqrt{N/t}$
simply by running algorithm $\textbf{Count}(f,L)$ with
exponentially increasing values of~$L$ until we obtain a non-zero
output. Having this rough estimate~$L$
of $\sqrt{N/t}$ we can then apply Theorem~\ref{thm:count} with $M$
in the order of~$\oneovereps L$ to find an
estimate~$\tilde{t}$ of~$t$ with the required accuracy. The precise
algorithm is as follows.

\begin{algorithm}{$\algapproxcount(f,\eps)$} \label{approx_count}
\begin{enumerate}
\item  Start with $\ell = 0$.
\item  Increase $\ell$ by~1.  \label{app_count}
\item  Set $t^{\prime} =$ \textbf{Count}$(f,2^\ell)$.
       \label{itemcheck2l}
\item  If $t^{\prime} = 0$ and $2^\ell < 2\sqrt{N}$ then
go to step~\ref{app_count}. \label{itemcheckupperbound}
\item Set $M= \big\lceil \frac{20 \pi^2}{\eps} 2^{\ell}\,\big\rceil$.
      \label{itemsetM}
\item \label{last_step}
 Set $t^{\prime} = \textbf{Count}(f,M)$.
 \item
  Output an integer $\tilde{t}$ satisfying
  $\left| \tilde{t} - t^{\prime} \right| \leq \frac{2}{3}$.
\end{enumerate}
\end{algorithm}

\begin{theorem}\label{thm:newrel}
Given a Boolean function $f$ with $N$ and $t$ defined as above,
and any $0 < \eps \leq 1$,
$\algapproxcount(f,\eps)$
outputs an estimate~$\tilde{t}$ such that
\begin{equation*}
\big|\tilde{t}-t\big| \; \leq \;
  \eps t
\end{equation*}
with probability
at least~$\frac23$,
using an expected number of
evaluations of~$f$ which is in
\mbox{$\Theta\big(\oneovereps\sqrt{N/t}\,\big)$}.
If $t=0$, the algorithm outputs $\tilde{t}=t$ with
certainty and $f$ is evaluated a number of times in $\Theta\big(\sqrt{N}\big)$.
\end{theorem}

\begin{proof}
When $t=0$, the analysis is straightforward.
For  $t>0$, let $\theta$ denote $\theta_{t/N}$ and
$m = \big\lfloor\mskip-2.2mu\log_2(\frac{1}{5\theta})\big\rfloor$.
{From} Theorem~\ref{QFT}
we have that the probability that step~\ref{itemcheck2l} outputs
{\bf Count}$(f,2^\ell) = 0$ for $\ell = 1,2,\ldots, m$ is
\begin{equation*}
   \prod^{m}_{\ell=1}  \frac{\sin^2(2^\ell \theta)}{2^{2\ell} \sin^2(\theta)}
   \,\geq\,
    \prod^{m}_{\ell=1} \cos^2(2^\ell \theta)
 \,=\,  \frac{\sin^2(2^{m+1}\theta)}{2^{2m}\sin^2(2 \theta)}
\,\geq\, \cos^2\big({\textstyle \frac{2}{5}}\big) \, .
\end{equation*}
The previous inequalities are obtained by using the fact that
$\sin(M\theta) \geq M \sin(\theta)\cos(M\theta)$ for any
$M \geq 0$ and $0 \leq M \theta < \frac{\pi}{2}$,
which can be readily seen by considering the Taylor expansion of
$\tan(x)$ at $x = M\theta$.

Now assuming step~\ref{itemcheck2l} has outputted $0$ at least $m$
times (note that \mbox{$2^m \leq \frac{1}{5 \theta} \leq
\frac{1}{5}\sqrt{N/t} < 2 \sqrt{N}$}),
after step~\ref{itemsetM} we have $M
\geq  \frac{20 \pi^2}{\eps} 2^{m+1} \geq
\frac{4 \pi^2}{\eps \theta}$
and by Theorem~\ref{thm:count}
(and the fact that $\theta \leq
\frac{\pi}{2} \sin(\theta) = \frac{\pi}{2} \sqrt{t/N}$)
the probability that {\bf Count}$(f, M)$ outputs an integer
$t^{\prime}$ satisfying $|t^{\prime} - t| \leq
\frac{\eps}{4} t+\frac{\eps^2}{64}t$
is at least~${8}/{\pi^2}$.
Let us suppose this is the case. If $\eps t < 1$, then
$|\tilde{t} - t| < 1$ and, since $\tilde{t}$ and $t$ are both
integers, we must have $t = \tilde{t}$.
If $\eps t \geq 1$,
then rounding off $t^{\prime}$ to $\tilde{t}$ introduces an error
of at most $\frac{2}{3} \leq \frac{2 \eps}{3} t$, making the
total error at most $\frac{\eps}{4} t+\frac{\eps^2}{64}t +
\frac{2 \eps}{3} t < \eps t$.
Therefore the overall probability of outputting an
estimate with error at most $\eps t$ is at least
$\cos^2\big( \frac{2}{5} \big) \times (8/\pi^2) > \frac{2}{3}$.

To~upper bound the number of applications of~$f$, note that by
Theorem~\ref{thm:count}, for any integer $L \geq 18 \pi
\sqrt{N/t}$, the probability that {\bf Count}$(f,L)$ outputs~0 is
less than~$1/4$. Thus the expected value of $M$ at
step~\ref{last_step} is in~$\Theta(\oneovereps \sqrt{N/t})$.
\end{proof}

We~remark that in algorithm \algapproxcount,
we could alternatively to steps 1 to~\ref{itemcheckupperbound}
use algorithm \algqsearch of Section~\ref{sec:ampl},
provided we have \algqsearch also output its final value of~$M$.
In~this case, we would use (a~multiple of) that value
as our rough estimate of~$\sqrt{N/t}$,
instead of using the final value of~$2^\ell$
found in step~\ref{itemcheckupperbound} of \algapproxcount.

Algorithm \algapproxcount is optimal for any fixed~$\eps$, but
not in general.  In Appendix \ref{approxcountproof} we give an
optimal algorithm, while we now present two simple optimal
algorithms for counting the number of solutions exactly. That is,
we now consider the problem of determining the exact value of $t =
|f^{-1}(-1)|$. In~the special case that we are given a nonzero
integer~$t_0$ and promised that either $t=0$ or $t=t_0$, then we
can determine which is the case with certainty using a number of
evaluations of $f$ in $O(\sqrt{N/t_0})$. This is an easy corollary
of Theorem~\ref{thm:witha} and we state it without proof.

\begin{theorem}\label{thm:promisedvaluesofa}
Let $f:\{0,1,\ldots,N-1\} \rightarrow \{0,1\}$ be a given Boolean function
such that the cardinality of the preimage of~1 is either~0 or~$t_0$.
Then there exists a quantum algorithm that determines with certainty
which is the case
using a number of evaluations of~$f$ which is in
$\Theta\big(\sqrt{N/t_0}\smallspace\big)$,
and in the latter case, also outputs a random element of $f^{-1}(1)$.
\end{theorem}

For the general case in which we do not have any prior
knowledge about~$t$, we offer the following algorithm.

\begin{algorithm}{$\algexactcount(f)$}\label{exact_count}
\begin{enumerate}
\item Set $t_1^{\prime}= {\textup{\textbf{Count}}}\big(f,
    \raisebox{\upbceil}{$\big\lceil$} 14 \pi \sqrt{N} \,
             \raisebox{\upbceil}{$\big\rceil$}\big)$
  and $t_2^{\prime}= {\textup{\textbf{Count}}}\big(f,
    \raisebox{\upbceil}{$\big\lceil$} 14 \pi \sqrt{N} \,
             \raisebox{\upbceil}{$\big\rceil$}\big)$.
\item Let $M_i = \raisebox{\upbceil}{$\big\lceil$} 30 \sqrt{(t_i^{\prime} +1)
             (N - t_i^{\prime} +1)} \,
           \raisebox{\upbceil}{$\big\rceil$}$
  for $i=1,2$.
\item Set $M = \min\{M_1, M_2\}$.
\item Set $t^{\prime} = \textbf{Count}(f,M)$.
\item
Output an integer $\tilde{t}$ satisfying $ \left| \tilde{t} -
t^{\prime} \right| \leq \frac{2}{3}$.
\end{enumerate}
\end{algorithm}


The main idea of this algorithm is the same as that of algorithm
\mbox{\algapproxcount}. First we find a rough estimate~$
t_r^{\prime}$ of~$t$, and then we run algorithm {\bf Count}$(f,M)$
with a value of~$M$ that depends on~$t_r^{\prime}$.
\mbox{By~Theorem~\ref{thm:count}}, if we set $M$ to be in the order of
$\sqrt{t_r^{\prime} (N - t_r^{\prime})}$, then the output
$t^{\prime}
= {\textbf{Count}}(f,M)$ is likely to be so that $|t^{\prime}
- t| < \frac{1}{3}$, in which case ${\tilde t} = t$.

\begin{theorem}\label{thm:exact}
Given a Boolean function $f$ with $N$ and $t$ defined as above,
algorithm
\algexactcount requires an expected number of
evaluations of $f$ which is in $\Theta(\sqrt{(t+1)(N-t+1)}\,)$ and outputs an
estimate~$\tilde{t}$ which equals $t$ with probability at
least~$\frac{2}{3}$ using space only linear in~$\log(N)$.
\end{theorem}

\begin{proof}
Apply Theorem~\ref{thm:count} with~$k=7$. For each $i=1,2$, with
probability greater than $\frac{11}{12}$, outcome ${t_i^{\prime}}$
satisfies $\big|{t_i^{\prime}} -t\big| <
\sqrt{\frac{t(N-t)}{N}} + 1/4$, in
which case we also have that $\sqrt{t(N-t)} \leq
\frac{\sqrt{2}}{30} M_i$. Thus, with probability greater than
$\left(\frac{11}{12} \right)^2$, we have
\begin{equation*}
\frac{\sqrt{t(N-t)}}{M} \;\leq\; \frac{\sqrt{2}}{30}.
\end{equation*}
Suppose this is the case.
Then by Theorem~\ref{thm:count},
with probability at least~$8/\pi^2$,
\begin{equation*}
|{t^{\prime}} - t|
\;\leq\;
     \frac{2 \pi \sqrt{2}}{30}
   + \frac{4 \pi^2}{30^2}
 \;<\; \frac{1}{3}
\end{equation*}
and consequently
\begin{equation*}
|{\tilde t} - t| < 1.
\end{equation*}
 Hence, with probability at least $\left( \frac{11}{12} \right)^2 \times
8/\pi^2
> \frac{2}{3}$, we have ${\tilde t} = t$.

The number of applications of~$f$ is
$ 2 \raisebox{\upbceil}{$\big\lceil$} 14 \pi \sqrt{N}\,
             \raisebox{\upbceil}{$\big\rceil$} +M$.
Consider the expected value of~$M_i$ for $i=1,2$.
Since
\begin{equation*}
  \sqrt{\vphantom{(N)}\smash{(t_i^{\prime}+1)(N-t_i^{\prime}+1)}}
  \;\leq\;
  \sqrt{(t+1)(N-t+1)}
  + \sqrt{\vphantom{(N)}\smash{N |t_i^{\prime} -t|}}
\end{equation*}
for any $0 \leq t_i^{\prime}, t \leq N$, we just need to upper
bound the expected value of~$\sqrt{N |t_i^{\prime} -t|}$. By
Theorem~\ref{thm:count}, for any $k\geq 2$,
\begin{equation*}
 | t_i^{\prime} - t |
\; \leq \;
  k \sqrt{\frac{t(N-t)}{N}} + k^2
\end{equation*}
with probability at least~$1 - \frac{1}{k}$. Hence $M_i$ is less than
\begin{equation}\label{eq:Msupperbound}
30(1+k) \left(  \sqrt{(t+1)(N-t+1)} + \sqrt{N} \,\right) + 1
\end{equation}
with probability at least~$1-\frac{1}{k}$.

In~particular, the minimum of $M_1$ and~$M_2$ is greater than the
expression given in Equation~\ref{eq:Msupperbound} with
probability at most~$\frac{1}{k^2}$. Since any positive random
variable~$Z$ satisfying $\textup{Prob}({Z} > k) \,\leq
\frac{1}{k^2}$ has expectation upper bounded by a constant, the
expected value of~$M$ is in~$O\big(\sqrt{(t+1)(N-t+1)}\,\big)$.
\end{proof}

It follows from Theorem~4.10 of~\cite{BBCMW}
that any quantum algorithm capable of deciding
with high probability whether or not a function
\mbox{$f:\{0,1,\ldots,N-1\}\rightarrow \{0,1\}$} is such that
\mbox{$| f^{-1}(1)| \leq t$}, given some \mbox{$0 < t < N$},
must query $f$ a number of times which is
at least in $\Omega\big(\sqrt{(t+1)(N-t+1)}\,\big)$ times.
Therefore, our exact counting algorithm is optimal up to
a constant factor.

Note also that successive applications of Grover's algorithm in
which we strike out the solutions as they are found will also
provide an algorithm to perform exact counting.  In order to obtain
a constant probability of success, if the algorithm fails to return
a new element, one must do more than a constant number of trials.
In particular, repeating until we get $\log(N)$ failures will
provide an overall constant probability of success.  Unfortunately,
the number of applications of~$f$
is then in $O\big(\sqrt{tN}+\log(N) \sqrt{N/t}\,\big)$
and the cost in terms of additional quantum memory is prohibitive,
that is in~$\Theta(t)$.

\section{Concluding remarks}\label{sec:conclusion}

Let $f: \{0,1,\ldots, N-1\} \rightarrow \{0,1\}$ be a function provided
as a black box, in the sense that the only knowledge available about $f$
is given by evaluating it on arbitrary points in its domain.
We~are interested in the number of times that $f$ must be evaluated
to achieve certain goals, and this number is our measure of efficiency.
Grover's algorithm can find the $x_0$ such that \mbox{$f(x_0)=1$}
quadratically faster in the expected sense than the best possible classical
algorithm provided the solution is known to be unique~\cite{Grover1,Grover2}.
We~have generalized Grover's algorithm in several directions.
\begin{itemize}
\itm The quadratic speedup remains when the solution is not unique,
even if the number of solutions is not known ahead of time.
\itm If the number of solutions is known (and nonzero), we can find one
quadratically faster in the worst case than would be possible classically
even in the expected case.
\itm If the number $t$ of solutions is known to be either 0 or~$t_0$, we can
tell which is the case with certainty, and exhibit a solution if \mbox{$t>0$},
in a time in $O(\sqrt{N/t_0}\,)$ in the worst case.
By~contrast, the best classical algorithm would need \mbox{$N-t_0+1$}
queries in the worst case.
This is much better than a quadratic speedup when $t_0$ is large.
\itm The quadratic speedup remains in a variety of settings that are not
constrained to the black-box model: even if additional information about $f$
can be used to design efficient classical heuristics, we can still find
solutions quadratically faster on a quantum computer, provided the heuristic
falls under the broad scope of our technique.
\itm We give efficient quantum algorithms to estimate the number of
\mbox{solutions} in a variety of error models.  In~all cases, our quantum
algo\-rithms are proven optimal, up to a multiplicative constant, among all
possible quantum algorithms.
In~most cases, our quantum algorithms are known to be quadratically
faster than the best possible classical algo\-rithm.
In~the case of counting
the number of solutions up to relative error~$\eps$, our optimal
quantum algorithm is quadratically faster than the best known classical
algorithm for fixed~$\eps$,
but in fact it is better than that when $\eps$ is not a constant.
Since we do not \mbox{believe} that a super-quadratic quantum improvement for
a non-promise black-box problem is possible,
we~conjecture that there exists a classical algorithm that uses
a number of queries in $O(\min\{M^2,N\})$,
where \mbox{$M = \sqrt{\frac{N}{\lfloor \eps t\rfloor +1}}
  + \frac{\sqrt{\vphantom{\tilde t}t(N-t)}}{\lfloor \eps t \rfloor +1}$}
is proportional to the number of queries \mbox{required} by our optimal
quantum algorithm.
This conjecture is further supported by the fact that we can easily
find a good estimate for~$M^2$, without prior knowledge of~$t$,
\mbox{using} a number of classical queries in $O(\oneovereps + \frac{N}{t+1})$.
\itm We can amplify efficiently the success probability not only of classical
search algorithms, but also of quantum algorithms.  More precisely, if a
quantum algorithm can output an $x$ that has probability \mbox{$a>0$} of being
such that \mbox{$f(x)=1$}, then a solution can be found after evaluating $f$
an expected number of time in $O(1/\sqrt{a}\,)$.
If~the value of $a$ is known, a solution can be found after evaluating $f$
a number of time in $O(1/\sqrt{a}\,)$ even in the worst case.
We~call this process \emph{amplitude amplification}.
Again, this is quadratically faster than would be possible
if the quantum search algorithm were available as a black box
to a classical algorithm.
\itm Finally, we provide a general technique, known as
\emph{amplitude estimation}, to estimate efficiently
the success probability $a$ of quantum search algorithms.
This is the natural quantum generalization of the above-mentioned technique
to estimate the number of classical solutions to the equation
\mbox{$f(x)=1$}.
\end{itemize}

The following table summarizes the
number of applications of the given function~$f$
in the quantum algorithms presented in this paper.
The table also compares the quantum complexities with the classical
complexities of these problems, when the latter are known.
Any lower bounds indicated (implicit in the use of the ``$\Theta$''
notation) correspond to those in the black-box model of computation.
In~the case of the efficiency of quantum counting with accuracy~$\eps$,
we refer to the algorithm given below in the Appendix.

\begin{center}
\begin{tabular}{|l|l|l|}
\hline
Problem & Quantum Complexity  & Classical Complexity  \\
\hline
Decision\strt
&  $\Theta{(\sqrt{N/(t+1)}\,)}$
&  $\Theta(N/(t+1))$ \\
Searching\strt
&  $\Theta{(\sqrt{N/(t+1)}\,)}$
&  $\Theta{(N/(t+1))}$  \\
Counting with error $\sqrt{t}$\strt
& $\Theta{(\sqrt{N}\,)}$
&   \\
Counting with accuracy~$\eps$\strt
& $\Theta\bigg(\sqrt{\frac{N}{\lfloor \eps t\rfloor +1}}
  + \frac{\sqrt{\vphantom{\tilde t}t(N-t)}}{\lfloor
                       \eps t \rfloor +1}\bigg)$
& $O(\frac{1}{\eps^2}{N/(t+1)})$  \\
Exact counting\strt
& $\Theta\big(\sqrt{(t+1)(N-t+1)}\,\big)$
&  $\Theta{(N)}$
\\
\hline
\end{tabular}
\end{center}

We leave as open the problem of finding a quantum algorithm that \mbox{exploits}
the structure of some searching or counting problem in a genuinely quantum way.
By~this, we mean in a way that is not equivalent to applying amplitude
amplification or amplitude estimation to a classical heuristic.  Note that
Shor's factoring algorithm does this in the different context of integer
factorization.

\section*{Acknowledgements}
We are grateful to Joan Boyar, Harry Buhrman, Artur Ekert, Ashwin
Nayak, Jeff Shallitt, Barbara Terhal and Ronald de~Wolf for helpful
discussions.

\appendix

\section{Tight Algorithm for Approximate\\Counting} \label{approxcountproof}

Here we combine the ideas of algorithms \algapproxcount and
\algexactcount to obtain an optimal algorithm for
approximately counting.  That this algorithm is optimal follows
readily from Corollary~1.2 and Theorem~1.13
of Nayak and Wu~\cite{NW}.

\begin{theorem}\label{thm:optappcount}
Given a Boolean function $f$ with $N$ and $t$ defined as above,
and any $\eps$ such that \mbox{$\frac{1}{3N} < \eps \leq 1$}, the following
algorithm $\algoptapproxcount(f,\eps)$
outputs an estimate~$\tilde{t}$ such that
\begin{equation*}
\big|\tilde{t}-t\big| \;\leq\; \eps t
\end{equation*}
with probability at least~$\frac23$,
using an expected number of evaluations of~$f$ in the order of
\begin{equation*}
\label{eq:optimalapprox}
S = \sqrt{\frac{N}{\lfloor \eps t\rfloor +1}}
  + \frac{\sqrt{\vphantom{\tilde t}t(N-t)}}{\lfloor
                       \eps t \rfloor +1}.
\end{equation*}
If~$t=0$ or $t=N$,
the algorithm outputs $\tilde{t}=t$ with certainty.
\end{theorem}

We~assume that $\eps N > 1/3$, since otherwise approximately
counting with accuracy~$\eps$ reduces to exact counting.
Set
\begin{equation}\label{eq:optapprox}
S' = \min \left\{\frac{1}{\sqrt{\eps}} \sqrt{\frac{N}{t}}
 \bigg( 1 + \sqrt{\frac{N-t}{\eps N}}\;\bigg)\,,
  \; \sqrt{(t+1)(N-t+1)} \,\right\}
\end{equation}
and note that $S' \in \Theta(S)$ where $S$ is defined
as in Theorem~\ref{thm:optappcount}.
The algorithm works by finding approximate values for
each of the different terms in Equation~\ref{eq:optapprox}.
The general outline of the algorithm is as follows.

\begin{algorithm}{$\algoptapproxcount(f,\eps)$}\label{alg:optappr}
\begin{enumerate}
\item Find integer $L_1$ approximating $\sqrt{N/(t+1)}$.
\item Find integer $L_2$ approximating $\sqrt{(N-t)/(\eps N)}$.
\item Set $M_1 = \frac{1}{\sqrt{\eps}} L_1 (1 + L_2)$.
      \label{item:setm1value}
\item If $M_1 > \sqrt{N}$ then find integer $M_2$
      approximating $\sqrt{(t+1)(N-t+1)}$.
      If $M_1 \leq \sqrt{N}$ then set $M_2 = \infty$.
      \label{item:m2approx}
\item Set $M = \min\{M_1,M_2\}$.
      \label{item:computeM}
\item Set $t^{\prime} = \textbf{Count}(f, \lceil 10 \pi M \rceil)$.
      \item Output an integer $\tilde{t}$ satisfying
            $\left| \tilde{t} - t^{\prime} \right| \leq \frac{2}{3}$.

\end{enumerate}
\end{algorithm}

\begin{proof}
To~find~$L_1$, we run steps 1 to~4 of algorithm
\algapproxcount and then
set $L_1 = \rblc{\upceil} 9 \pi \times 2^l \rbrc{\upceil}$.
A~proof analogous to that of Theorem~\ref{thm:newrel} gives that
\begin{itemize}
\item $L_1 > \sqrt{N/(t+1)}$ with probability at least~$0.95$, and
\item the expected value of $L_1$ is in $\Theta\big(\sqrt{N/(t+1)}\,\big)$.
\end{itemize}
This requires a number of evaluations of~$f$ which is in $\Theta(L_1)$ ,
and thus,  the expected number of evaluations of~$f$ so far is in~$O(S')$.

In~step~2, for some constant $c$ to be determined below,
we use $2 \big\lceil\frac{c}{\sqrt{\eps}}\big\rceil$
evaluations of~$f$ to find integer~$L_2$
satisfying
\begin{itemize}
\item $L_2 > \sqrt{(N-t)/(\eps N)}$ with probability
at least~$0.95$, and
\item the expected value of $L_2$
is in $O\big(\sqrt{(N-t+1)/(\eps N)}\,\big)$.
\end{itemize}
Since $N-t = |f^{-1}(0)|$, finding such $L_2$ boils down to estimating,
with accuracy in~$\Theta(\sqrt{\eps}\,)$,
the square root of the probability that $f$ takes
the value~0 on a random point in its domain.
Or~equivalently, the probability that
$\neg f$ takes the value~1, where $\neg f = 1-f$.
Suppose for some constant $c$,
we run
$\algcount(\neg f, \big\lceil\frac{c}{\sqrt{\eps}}\big\rceil)$
twice with outputs ${\tilde r}_1$ and~${\tilde r}_2$.
By~Theorem~\ref{thm:count}, each output~${\tilde r}_i$
($i=1,2$) satisfies that
\begin{equation*}
  \left| \sqrt{\frac{{\tilde r}_i}{\eps N}}
   - \sqrt{\frac{N-t}{\eps N}} \,\right|
  \;\leq\; \sqrt{\frac{2 \pi k}{c}}
   \,\sqrt[{\textnormal{\footnotesize{4}}}]{\frac{N-t}{\eps N}}
  + \frac{\pi k}{c}
\end{equation*}
with probability at least $1 - \frac{1}{2(k-1)}$
for every $k \geq 2$.
It~follows that
${\tilde r} = \min\big\{\sqrt{{\tilde r}_1/(\eps N)},
  \sqrt{{\tilde r}_2/(\eps N)}\big\}$
has expected value in $O\big(\sqrt{(N-t+1)/(\eps N)}\,\big)$.
Setting $k=21$, $c = 8 \pi k$, and $L_2 = \lceil 2 {\tilde r}
\rceil +1$, ensures that $L_2$ satisfies the two properties
mentioned above.
The number of evaluations of~$f$ in step~2
is in $\Theta(\frac{1}{\sqrt{\eps}})$ which is in~$O(S')$.

In~step~\ref{item:setm1value}, we set
$M_1 = \frac{1}{\sqrt{\eps}} L_1 (1 + L_2)$.
Note that
\begin{itemize}
\item $M_1 > \frac{1}{\sqrt{\eps}} \,\sqrt{\frac{N}{t+1}}
       \left( 1+ \sqrt{\frac{N-t}{\eps N}} \,\right)$
      with probability at least~$0.95^2$, and
\item the expected value of $M_1$ is in the order of
      $\frac{1}{\sqrt{\eps}} \,\sqrt{\frac{N}{t+1}}
       \left( 1+ \sqrt{\frac{N-t+1}{\eps N}} \,\right)$.
\end{itemize}

In~step~\ref{item:m2approx}, analogously to algorithm
\algexactcount, a number of evaluations of $f$ in
$\Theta(\sqrt{N})$ suffices to find an integer~$M_2$ such that
\begin{itemize}
\item $M_2 > \sqrt{(t+1)(N-t+1)}$
      with probability at least~$0.95$, and
\item the expected value of $M_2$
      is in $\Theta\big(\sqrt{(t+1)(N-t+1)}\,\big)$.
\end{itemize}
Fortunately, since $\sqrt{(t+1)(N-t+1)} \geq \sqrt{N}$, we shall
only need~$M_2$ if $M_1 > \sqrt{N}$. We~obtain that, after step~5,
\begin{itemize}
\item $M$ is greater than
\begin{equation*}
       \min\left\{
       \frac{1}{\sqrt{\eps}} \,\sqrt{\frac{N}{t+1}}
       \left( 1+ \sqrt{\frac{N-t}{\eps N}} \,\right),
       \;
       \sqrt{(t+1)(N-t+1)} \right\}
\end{equation*}
      with probability at least~$0.95^3 > 0.85$, and
\item the expected value of $M$ is in~$O(S')$.
\end{itemize}
To~derive this latter statement,
we use the fact that the expected value of
the minimum of two random variables is at most
the minimum of their expectation.

Finally, by Theorem~\ref{thm:count}, applying algorithm
$\algcount(f, \lceil 10 \pi M \rceil)$ given such an~$M$, produces
an estimate~$t^{\prime}$ of~$t$ such that $|t^{\prime} -t| \leq
\frac{\eps t}{3}$
(which implies that
$|\tilde{t} - t | \leq \eps t$) with probability at
least~$8/\pi^2$. Hence our overall success probability is at least
$0.85 \times 8/\pi^2 > 2/3$, and the expected number of evaluations
of~$f$ is in~$O(S')$.
\end{proof}


\pagebreak[1]

\end{document}